\documentclass{aa501}
\usepackage{graphicx}

\begin{document}
\input{psfig.sty}

\title{
Extragalactic Globular Clusters in the Near Infrared {\tt I}:\\
A comparison between M87 and NGC 4478 
\thanks{Data presented herein were obtained at the W.M. Keck Observatory, 
which is operated as a scientific partnership among the California Institute 
of Technology, the University of California and the National Aeronautics and 
Space Administration. The Observatory was made possible by the generous 
financial support of the W.M. Keck Foundation.}
}

\author {Markus Kissler-Patig
	\inst{1}
	\and
	Jean P. Brodie
	\inst{2}
	\and
	Dante Minniti
	\inst{3} }

\offprints {Markus Kissler-Patig \email{mkissler@eso.org}}

\institute{European Southern Observatory, Karl-Schwarzschild-Str.~2, D-85748
Garching, Germany \\  \email{mkissler@eso.org}
\and
UCO/Lick observatory, University of California, Santa Cruz, CA 95064, USA \\
\email{brodie@ucolick.org}
\and
Departamento de Astronom\'\i a y Astrof\'\i sica, P.~Universidad
Cat\'olica, Casilla 104, Santiago 22, Chile \\ \email{dante@astro.puc.cl}
}

\date{Received June 30, 2001; accepted July 15, 2001}

\abstract{ We compare optical and near infrared colours of globular
clusters in M87, the central giant elliptical in Virgo, and NGC 4478, an
intermediate luminosity galaxy in Virgo, close in projection to M87.  Combining $V$ and $I$ photometry obtained with the WFPC2 on HST and $K_s$
photometry obtained with the NIRC on Keck 1, we find the broad range
in colour and previously detected bi-modality in M87. We confirm that NGC 4478
only hosts a blue sub-population of globular clusters and now show that
these clusters' $V-I$ and $V-K$ colours are very similar to those of the 
halo globular clusters in
Milky Way and M31. Most likely, a metal-rich sub-population never formed
around this galaxy (rather than having formed and been destroyed later),
probably because its metal-rich gas was stripped during its passage through the
centre of the Virgo cluster.\\ The $V-I$, $V-K$ colours are close to
the predicted colours from SSP models for old populations. However, M87
hosts a few red clusters that are best explained by intermediate ages (a few
Gyr). Generally, there is evidence that the red, metal-rich sub-population
has a complex colour structure and is itself composed of clusters spanning a 
large metallicity and, potentially, age range. This contrasts with the blue,
metal-poor population which appears very homogeneous in all galaxies
observed so far.   }

\authorrunning{Kissler-Patig et al.}
\titlerunning{Extragalactic Globular Clusters in the Near Infrared {\tt I}}
\maketitle


\section{Introduction}

Globular cluster systems are now established as useful tools for
studying the formation and evolution of their host galaxies (see
recent reviews by Ashman \& Zepf 1998, Kissler-Patig 2000, Harris
2001).  One of the most interesting aspects of this approach is the
composite nature of most globular cluster systems. Many systems of
early-type galaxies appear to be composed of two or more distinct
sub-populations of globular clusters (Zepf \& Ashman 1993, see also
recently Gebhardt \& Kissler-Patig 1999, Larsen et al.~2001, and Kundu
\& Whitmore 2001a,b).  A number of groups have concentrated on
studying the properties of the individual globular cluster
sub-populations in a galaxy (e.g.~Kissler-Patig 2002 for a recent
overview, or above reviews) with the goal of revealing the origins of
the sub-populations and thus shedding light on the host galaxy
formation and evolution. To summarize the results so far, the blue and
the red globular cluster sub-populations differ significantly in their
spatial distributions, mean sizes, metallicities and kinematics while,
in contrast, they appear to have similar old ages within the large
(2--4 Gyr) errors.  All this indicates that there must have been
significant differences in the origins of the two main sub-populations
(e.g.~Burgarella et al.~2001 and references therein). 

The keys to studying differences in sub-population properties is their
unambiguous identification, and a reliable
association of the individual clusters with the different sub-populations.
The identification of sub-populations started in the Milky Way with the 
landmark paper of Zinn (1985), but another decade elapsed before they were
revealed in
early-type galaxies (Zepf \& Ashman 1993, Lee \& Geisler 1993, 
and above references for the long lists of following studies).
Observationally, the sub-populations are primarily identified in the 
metallicity distribution of the entire system. While for nearby
galaxies (including the Milky Way)
the metallicities are determined spectroscopically, we rely on
photometric studies and colour-to-metallicity conversions for the vast
majority of (further) early-type galaxies.

This immediately prompts the questions: how metallicity sensitive is the
photometry? Especially given the age-metallicity degeneracy of broad
band colours (e.g.~Worthey 1994). 
Most studies (in particular the ones done with the WFPC2) used broad
band colours in the Johnson-Cousins systems or equivalent which are
poorly suited to disentangling sub-populations.

This motivated our choice to work with a combination of optical and
near-infrared colours. The long wavelength baseline provides the most
metallicity sensitivity, allowing the sub-populations to be distinguished
by their colours even if they are very close in metallicity.
Historically, the first near-infrared measurements of extragalactic
globular clusters were carried out in M31 (Frogel, Persson \& Cohen 1980) 
and the Large Magellanic Cloud (Persson et al.~1983).
Technical advancement both in the near-infrared arrays, as well as in
telescope sizes allows us now to study globular clusters beyond the
Local Group with high photometric accuracy. The first studies beyond the
Local Group focused on young/intermediate age star clusters, whose AGB
stars contribute significantly to the K magnitude (see Minniti et
al.~1996 and Rejkuba 2001 for studies in Cen A, Maraston et
al.~2001 for studies in NGC 7252, Goudfrooij et al.~2001 for studies of 
NGC 1316). 

Our larger project aims at studying a dozen galaxies of different
luminosities and in different environments. In the first paper of the
series, we study the Virgo Galaxies Messier 87 and NGC~4478. Their
properties are summarized in Table \ref{t:galprop}.
M87 is the central giant elliptical in Virgo while NGC 4478 is an
intermediate luminosity galaxy in Virgo, close in projection to M87. 

\begin{table*}[ht]
\centering
\caption{Some properties of M~87 and NGC~4478}
\label{t:galprop}
\begin{tabular}{l r r l}
\hline
\noalign{\smallskip}
Property & M~87 & NGC~4478 & Reference\\
\noalign{\smallskip}
\hline
\noalign{\smallskip}
RA(2000) & 12 30 49.7 & 12 30 17.4& RC3 (de Vaucouleurs et al.~1991)\\
DEC(2000)& $+$12 23 24 & $+$12 19 44& RC3\\
B$_T{}_0$ & 9.49 & 12.21 & RC3\\
$(B-V)_0$ & 0.93 & 0.88 & RC3\\
$(m-M)$ & 31.03$\pm0.16$ & 31.29$\pm 0.28$ & Tonry et al.~(2000)\\
M$_V$ & $-22.47\pm0.18$ & $-19.96\pm0.30$ & derived from the above data\\
\noalign{\smallskip}
\hline
\end{tabular}
\end{table*}

Our near-infrared data (NIR) and the combination with optical data from
WFPC2 on HST, are described in Sect.~2. The $V,I,K$ colour distributions
presented in Sect.~3. Our goals are threefold: First, we will study the
colour / metallicity distribution of the globular clusters systems, taking
advantage of the higher metallicity resolution afforded by the NIR, to shed
light on the sub-populations origins.  In particular, NGC 4478 has been
reported as having hardly any metal-rich globular clusters (Neilsen,
Tsvetanov \& Ford 1997), which is unusual for a galaxy of that
luminosity. Second, we will compare the optical -- near-infrared colours
with existing population synthesis models and data from the Milky Way and
M31. This is valuable because NIR data of ``clean'' single stellar
populations with high metallicities, suitable for comparison with models,
are sparse.  Finally, for M87, we will look for sub-structure within the red
sub-population and probe the metallicity dispersion of the metal-rich
clusters.
A summary of our findings and conclusions are given in Sect.~5.



\begin{figure}[ht]
\psfig{figure=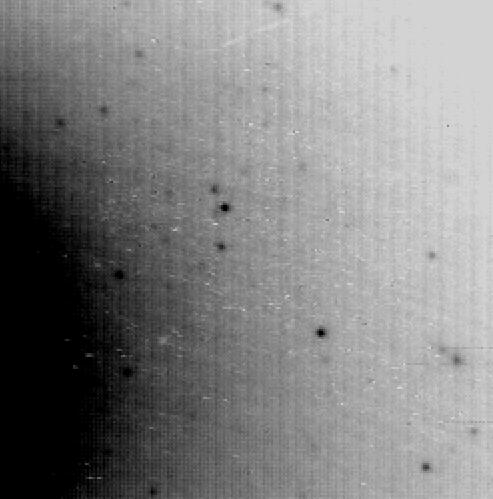,height=7cm
,bbllx=8mm,bblly=57mm,bburx=205mm,bbury=245mm}
\psfig{figure=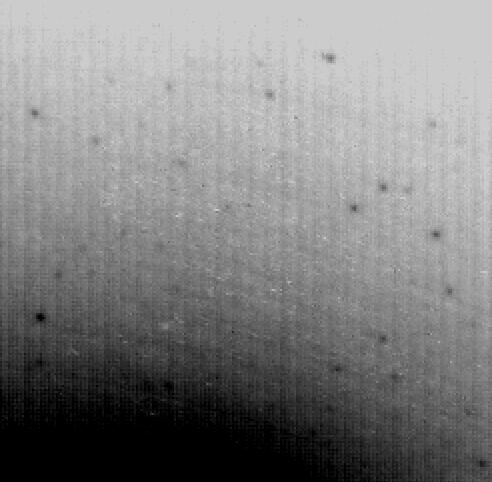,height=7cm
,bbllx=8mm,bblly=57mm,bburx=205mm,bbury=245mm}
\caption{ 
Images of our two M87 fields. Upper panel: the M87 `east'
field, $K_s$ filter, 650s exposure; lower panel: the M87 `west' field, $K_s$ 
filter, 750s exposure.  
Both images are 38\arcsec$\times$38\arcsec in size, for the orientation
see Fig.~\ref{fig:redpos}.
}
\label{fig:imm87}
\end {figure}

\begin{figure}[ht]
\psfig{figure=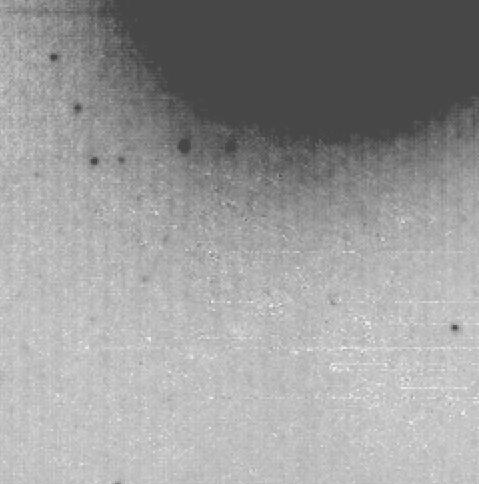,height=7cm
,bbllx=8mm,bblly=57mm,bburx=205mm,bbury=245mm}
\psfig{figure=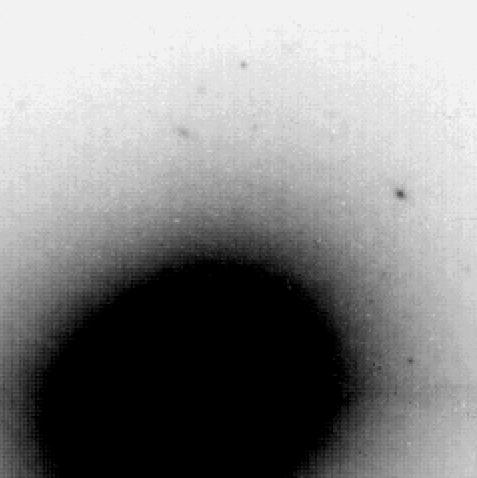,height=7cm
,bbllx=8mm,bblly=57mm,bburx=205mm,bbury=245mm}
\caption{ 
Images of our two NGC 4478 fields. Upper panel: the NGC 4478 `north'
field, $K_s$ filter, 3300s exposure; lower panel: the NGC 4478 `south' field, 
$K_s$ filter, 3300s exposure. 
Both images are 38\arcsec$\times$38\arcsec in size. Same orientation on
the sky as for the M87 images.
}
\label{fig:imn4478}
\end {figure}


\section{Observations and Reductions}


\begin{figure}[ht]
\psfig{figure=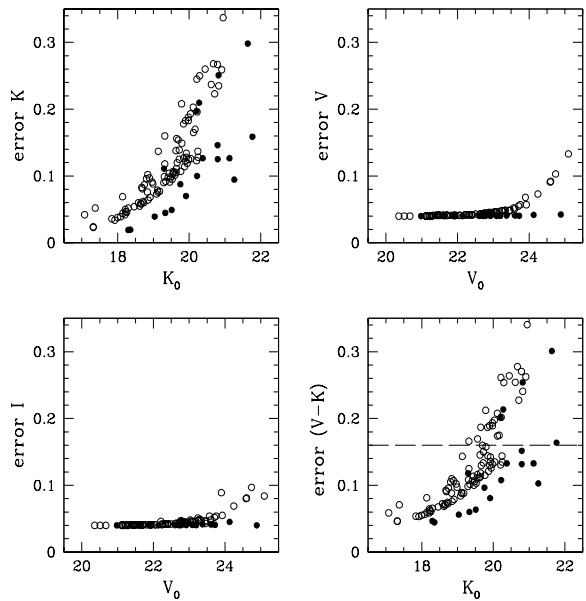,height=8cm,width=8cm
,bbllx=8mm,bblly=57mm,bburx=205mm,bbury=245mm}
\caption{ 
Error in $K,V,I,(V-K)$ plotted as a function of $K$ and $V$ magnitude
for globular cluster candidates in M87 (open circles) and NGC 4478 
(solid dots). The dashed line in the lower right panel indicates our error 
limit (0.16 mag) used later in the paper. Note that the error in
$(V-K)$ is completely dominated by the error in $K$.
}
\label{fig:errors}
\end {figure}


\subsection{Near Infrared Photometry}

The near infrared photometry was obtained on the night of March 6th,
1998 with the Near Infrared Camera (NIRC, Matthews \& Soifer 1994) at
the Keck 1 telescope. The camera is equipped with a $256 \times 256$
InSb array mounted at the f/25 forward Cassegrain focus of the
telescope; a 30 $\mu$m pixel corresponds to 0.15\arcsec~on the sky; the
total field of view is $38.4\arcsec \times 38.4\arcsec$. The
observations were made using the $K_s$ (1.99$\mu$m -- 2.32$\mu$m)
filter. 

We selected our NIRC fields to lie within the available WFPC2 images in
order to have optical colours for all our target globular clusters.  We
observed two globular cluster fields in M87, adjacent (East--West),
40\arcsec~south of the centre, with repeated observations of a sky field,
several arcminutes away.  In NGC 4478, we observed two globular cluster
fields covering the centre of the galaxy (one to the North and one to the
South), as well as a field $\simeq 2\arcmin$ away that served as sky field
but was still included in the WFPC2 field of view.

The total integration times for the different fields were: 750s for the
M87 west field, 650s for the M87 east field, and 3300s for each of the three
fields around NGC 4478. The seeing on all exposures, as measured from
the FWHM of point sources, varied between 0.4\arcsec~and 0.6\arcsec.
A typical observing cycle of $\simeq 5$ minutes included the two target
fields and the sky field. The sky exposures were 
dithered in a box pattern with a total integration time equivalent to the 
target exposures. A median of the sky exposures was then subtracted from
the target exposures to correct for dark current and sky variations. A
master flat-field, obtained from all sky exposures of the night, was
further used to correct for pixel to pixel variations. The
target exposures thus corrected were then averaged to form a final image.

The galaxy light was efficiently removed from the frames by modelling
it with a $35\times35$ pixel median filter. When the galaxy centre was
located in the field, i.e.~for the NGC 4478 south field, an isophotal
model was computed with {\sc isophote} under {\sc iraf}. These models
were subtracted from the original frames before the final object
detection was started.  The object finding and aperture photometry was
carried out with {\sc sextractor} (Bertin \& Arnouts 1996). The fields
are sufficiently un-crowded that large apertures (10--15 pixels
radius) could be used and total magnitudes could be derived from a
curve of growth analysis. The errors in the photometry were derived
from {\sc sextractor} and are shown in Fig.~\ref{fig:errors}. They are
dominated by photon noise (both of the sky background and the galaxy
background) as verified by measuring the bright objects on individual
exposures, with different sky subtractions, and by measuring all
objects on the flat-fielded and non flat-fielded images (checking for
systematic errors), as well as in apertures of different
sizes. Fig.\ref{fig:errors} (upper left panel) shows that for a given
$K$ magnitude the error varies significantly according to the position
of the object with respect to the galaxy background. Also it becomes
clear that the error in $(V-K)$ is completely dominated by the error
in $K$.  Only around $K\sim20$ ($V\sim23$) does the error in $V$ start
to noticeably contribute and to steepen the slope with which the error
in $(V-K)$ varies with $K$ (cf.~lower right panel). In addition, a
group of red objects close to the galaxy center (on high background
noise) mimic a ``jump'' in the error around $K\sim20$.  For all these
reasons we chose, in what follows, to consider only clusters with
$(V-K)$ errors less than 0.16 mag.

The calibration was performed with 6 standard stars (of which one was observed
twice) from Persson et al.'s (1998) HST $JHK$
standard list (SJ9108, SJ9110, SJ9134, SJ9143, SJ9149 (observed twice), and
SJ9155). The standards were distributed over the night; each standard
was observed in a $3\times3$ mosaic (medianed to compute the sky) and 
magnitudes were measured from each of the 9 sky--subtracted images. 
The 63 standard star measurements were used to derive a zero point and
extinction
coefficient; the colour term was assumed to be negligible. We obtained
the following relation:\\ $m_{cal}({\rm K}_s) = m_{inst} + 24.847 (\pm0.014)  -
0.045 (\pm 0.005) \cdot AM$,\\ where $m_{inst}$ is the measured
magnitude of an object (normalized to 1s exposure), $AM$ is the airmass,
and $m_{cal}({\rm K}_s)$ the calibrated magnitude we are seeking.

\subsection{Optical photometry}

The optical photometry was obtained from WFPC2 images of the HST Archive. 
For M87, we used the observations from proposal GO-5477 described in Whitmore
et al.~(1995). These images were obtained in the F555W ($4 \times 600$s) and F814W 
($4 \times 600$s) filters.
For NGC 4478, we used HST parallel observations with WFPC2 described
in Neilsen et al.~(1997). Images were obtained in the F606W ($6\times
2800$s) and the F814W ($5\times 2800$s $+ 2500$s).

Both Whitmore et al.~(1995) and Neilsen et al.~(1997) comprehensively
investigated the globular cluster systems of the host galaxies in $V$ and $I$.
Kundu et al.~(1999) and Larsen et al.~(2001) have since re-investigated 
the M87 data.
We will come back to some of their results in the next sections.
We chose to reduce the data independently, following the procedure
described in Puzia et al.~(1999). In summary, the images were combined
and cosmic ray cleaned within {\sc IRAF}, and object magnitudes measured in 2
pixels apertures using {\sc Sextractor}. Aperture correction derived in
Puzia et al.~(1999) for globular clusters in NGC~4472 (a Virgo galaxy
at a similar distance to M87 and NGC~4478) were applied to
correct for the extended nature of the clusters. The standard
calibrations given in Holtzman et al.~(1995) were used to compute Johnson 
$V$ and $I$ magnitudes. The results were compared with the lists of Whitmore et
al.~and Neilsen et al.~, and found to be in excellent agreement. 

Globular cluster candidates were selected from the HST data using the image
shape parameters returned by {\sc sextractor}. The optical and
near-infrared lists were combined to final candidate lists, including 93
objects in M87 and 19 in NGC 4478. Their magnitudes are plotted against
the total $V-K$ errors in Fig.~\ref{fig:errors}. The limiting magnitudes
for detection are roughly $K\sim21$ and $K\sim22$ for M87 and NGC 4478,
respectively. For some later applications, we will restrict the sample
to clusters with errors in $(V-K)<0.16$ mag (see above), leaving 70 clusters 
in M87 and 14 in NGC 4478, and reducing the limiting magnitudes by $\sim 1$
mag in each case.

\subsection{Spectroscopy}

For M87, we checked the literature to see if any of the clusters for which we obtained
a $K$-band magnitude had been included in a spectroscopic sample. 
Unfortunately, this was not the case, mainly due to
the proximity of our fields to the centre of M87. On the other hand,
this underscores an advantage of NIR studies that can probe, with high
metallicity sensitivity, globular clusters in the inner regions of
galaxies, inaccessible to spectroscopic studies.


\section{The $V,I,K$ Colours of the Globular Clusters}

\begin{figure}[ht]
\psfig{figure=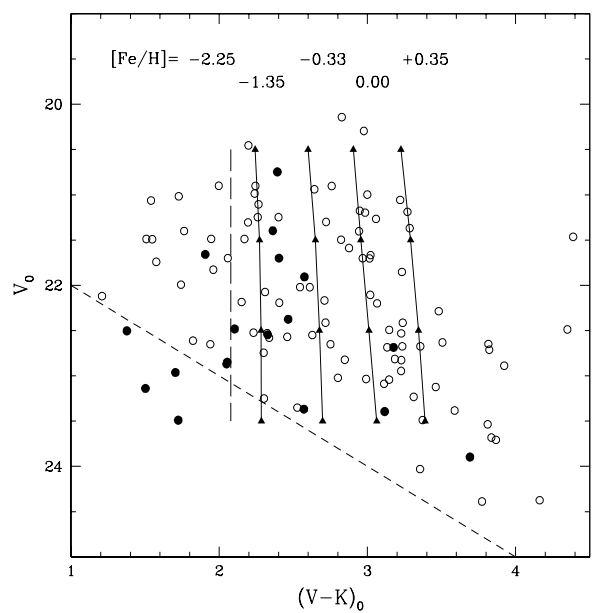,height=8cm,width=8cm
,bbllx=8mm,bblly=57mm,bburx=205mm,bbury=245mm}
\caption{ 
Colour magnitude diagrams for globular clusters in M87 (open circles),
and NGC~4478 (solid dots). The short dashed line indicates a line of equal 
detection probability (arbitrarily shifted here to $K=21$ which roughly
corresponds the $(V-K)$ error limited adopted throughout the paper). For five
metallicities, we show the predicted {\it colours} from simple stellar
population models (here using Maraston 2000) for four ages (15, 13, 11,
and 9 Gyr, from top to bottom, except for
[Fe/H]$=-2.25$ for which only models for 15 Gyr were available). The
models points are arbitrarily shifted in magnitude (driven primarily
by the mass and not the age of the clusters).
}
\label{fig:cmd}
\end {figure}

The colour-magnitude diagrams ($V$ against $V-K$) for the globular cluster in 
M87 and NGC~4478 are shown in Fig.~\ref{fig:cmd}. We over-plotted models from
Maraston (2000) as metallicity (age) references, and a line showing our
equal detection probability: the data are complete down to $V>24$, however
only to $K\leq 21$ and $K\leq 22$ for M87 and NGC~4478, respectively.
Thus, we detect red clusters more easily than blue ones.

Throughout the paper we will compare the data with population synthesis
models. We use models from Maraston (2000) and Bruzual \& Charlot
(2000) to illustrate the differences between current models in the NIR. 
For detailed descriptions we refer the reader to the original papers,
listing the main differences briefly below.
Both models use the same model atmospheres (``Basel atmospheres'',
Lejeune, Buser \& Cuisinier 1998), but differ in the stellar evolution tracks:
while Maraston uses tracks from the Frascati database (Cassisi et
al.~1999 and references therein), Bruzual \& Charlot use the Padova
tracks (Alongi et al.~1993, Girardi et al.~1996 and references therein). 
Both sets of tracks are overall very similar but differ in the
slope of the AGB. Further, the main difference between the models is
the fact that Maraston computes her models using the fuel consumption method 
(Renzini \& Buzzoni 1986) while Bruzual \& Charlot use the more classical 
isochrone synthesis method. The two last points taken together slightly 
affect the $K$ and $V-K$ predictions at high 
metallicities, as well as significantly affecting the isochrones for young 
ages ($<1 Gyr$, see Maraston et al.~2000). The latter is, however, of no 
concern for this paper. The difference between the models is well illustrated 
in Fig.~\ref{f:m87_models}, where for $V-K>3$ the 15 Gyr Bruzual \&
Charlot isochrone overlaps with the 9 Gyr isochrone of Maraston.
However, for ages younger than $\sim 9$ Gyr and high metallicities, both 
models predict the same trend. Thus the model uncertainties prevent us 
from reliably assigning {\it absolute} ages, but the choice of a particular 
model will not significantly affect our overall results. 

\begin{figure}[ht]
\psfig{figure=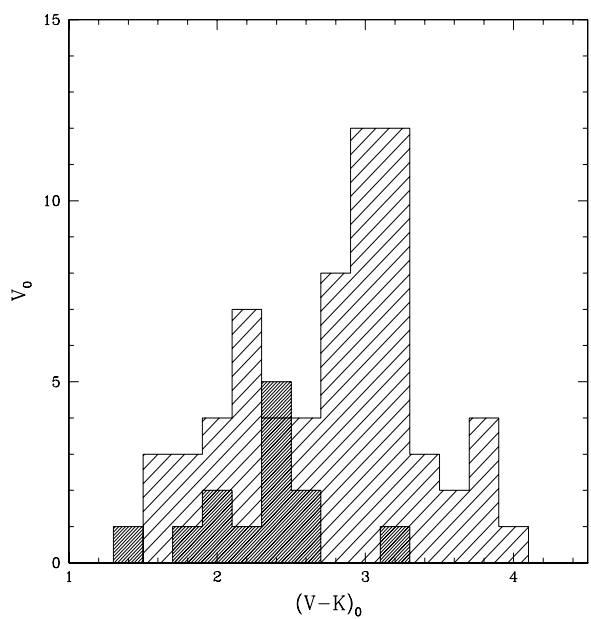,height=8cm,width=8cm
,bbllx=8mm,bblly=57mm,bburx=205mm,bbury=245mm}
\caption{ 
$V-K$ histogram for M87 (light hashed) and NGC~4478 (dark hashed)
globular clusters. Note that the histograms are biased in favour of red
clusters (see text).
}
\label{fig:vkhisto}
\end {figure}

The colour histograms are shown in Fig.~\ref{fig:vkhisto}. As
indicated in Fig.~\ref{fig:cmd}, our observations are biased in favour
of red clusters. Quantitatively, the following corrections should be
applied. The peak magnitude of the Gaussian globular cluster
luminosity function lies around $V_{\rm TO}\sim23.8$ at the Virgo
distance (e.g.~Puzia et al.~1999), and is similar to within a few
tenths of a magnitude for blue and red clusters. Thus, we expect the
peaks in $K$ to lie around $K_{\rm TO}\sim21.6$ for blue
($V-K\sim2.2$) clusters, and $K_{\rm TO}\sim20.7$ for red
($V-K\sim3.2$) clusters. The $V$ band luminosity function has a width
of $\sigma = 1.0$ to 1.2 magnitudes, and is expected to be similar in
$K$ for clusters in a narrow colour range.

For M87, our limiting
$K$ magnitude lies around 21 mag, so that for clusters with $V-K \sim 2.2$
we fall short of the peak of the luminosity function by $\sim 0.5
\sigma$, i.e.~sampling only around 30\% of the blue population in the area covered. For red clusters, however, our data are deep enough to pass the peak of
the luminosity function by $\sim 0.3\sigma$, i.e.~we sample around 60\%
of the red population in the area covered. Thus, in the case of M87,
the blue side of the histogram should be corrected by roughly a factor
of 2 when compared to the red side.

The effect is similar for NGC~4478. However, since our limiting
magnitude lies around $K\sim22$, in the area covered, we sample around 60\% and 85\% of the
blue and red population, respectively. Thus the
correction to apply when comparing blue and red numbers is smaller.

\begin{figure}[ht]
\psfig{figure=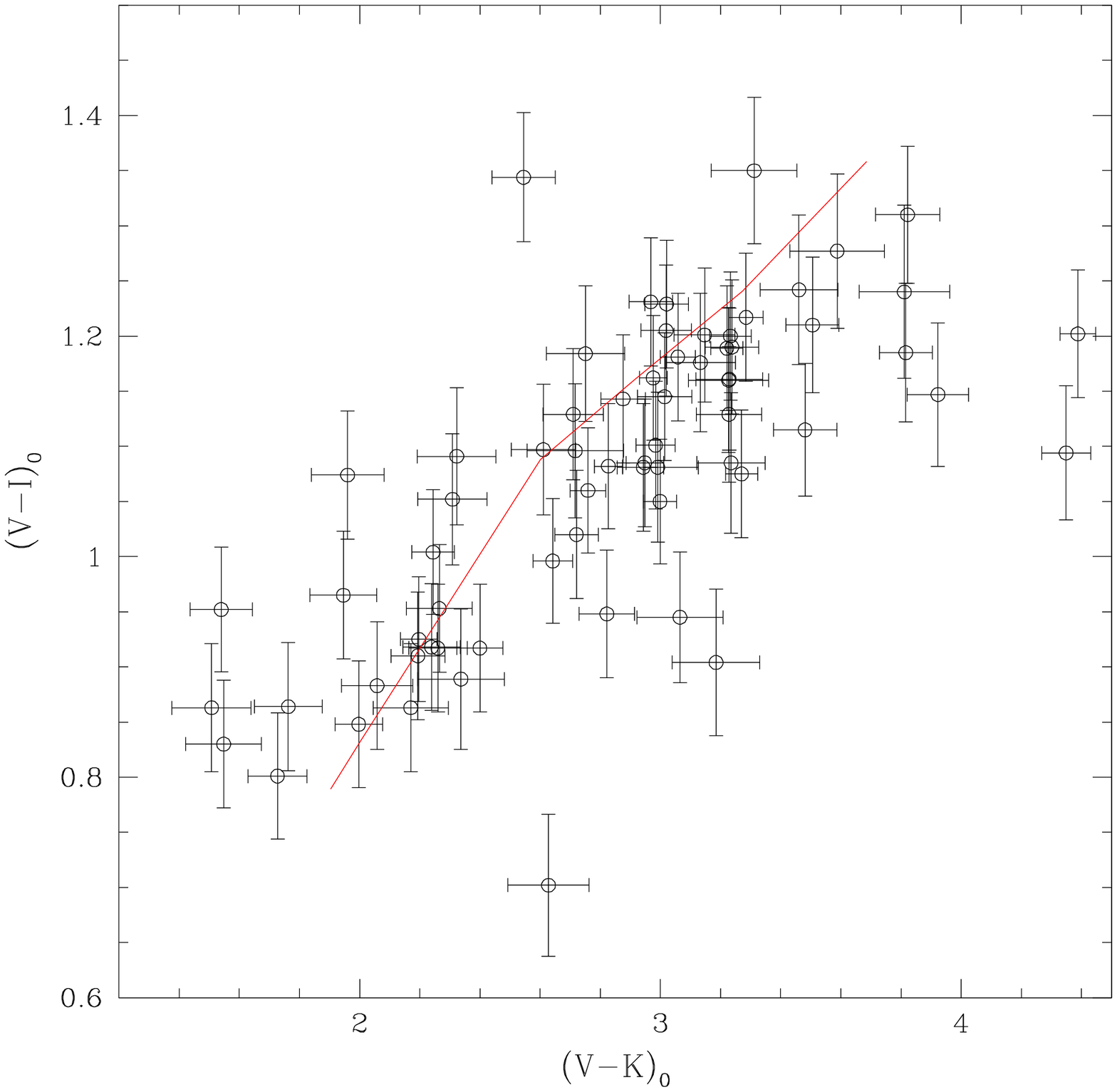,height=8cm,width=8cm
,bbllx=8mm,bblly=57mm,bburx=205mm,bbury=245mm}
\psfig{figure=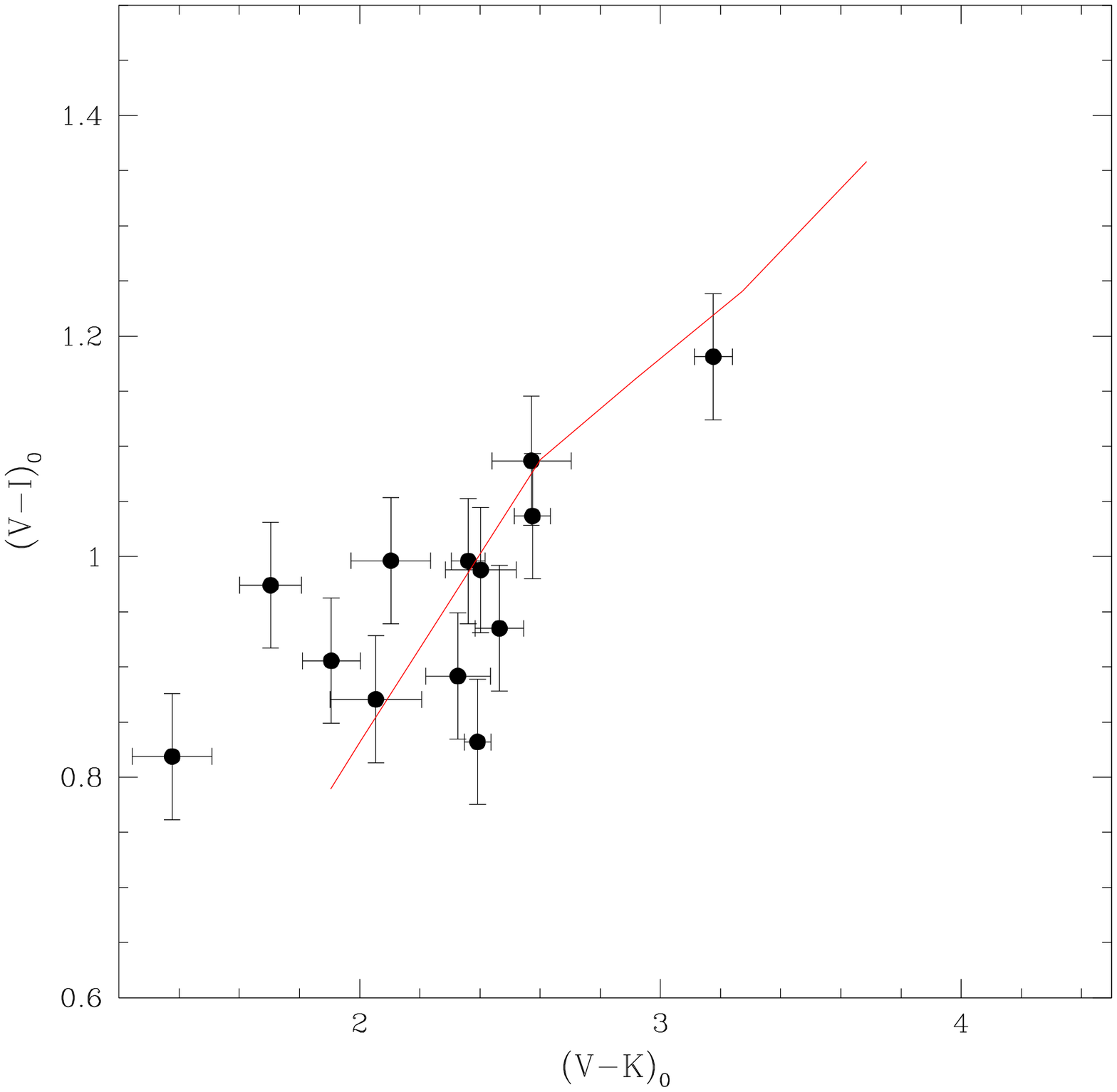,height=8cm,width=8cm
,bbllx=8mm,bblly=57mm,bburx=205mm,bbury=245mm}
\caption{ 
Colour-colour diagram for globular clusters in M87 (upper panel) and
NGC~4478 (lower panel). Only clusters with errors in $V-K$ less
than 0.16 mag are shown (all have errors in $V-I<0.1$). 
For comparison, a solid line shows a 15 Gyr isochrone
(from $1/200$ to twice solar metallicity) of a simple stellar population
(using here Bruzual \& Charlot (2000) models).
}
\label{f:vivk}
\end {figure}

Finally, we show in Fig.~\ref{f:vivk} the $V-I$ against $V-K$ colours for
the clusters, including the errors. Note that despite the fact that the 
$V-K$ photometric
errors are larger than the $V-I$ ones, the ratio of the errors with respect
to the colour range spanned is more favourable for $V-K$ by a factor of
about 2. Indeed, the mean error in
$V-K$ is 0.098 mag for this sample (0.129 when the sample is not restricted
to clusters with
errors $<0.16$) for a range in colour of $>1.5$ mag. The mean
error in $V-I$ is 0.061 mag (0.066 mag for the unrestricted sample) and the
range in colour is $\sim 0.5$. Thus, the colour ``resolution'' in $V-K$ is
$1.5/0.098 \simeq 15$, while in $V-I$ it is $0.5/0.061 \simeq 8$.

We will show further versions of Fig.~\ref{f:vivk}
when comparing our data to Milky Way and M31 clusters, as well as
to population synthesis models in Sect.~\ref{s:comp}.

\subsection{Transforming (V-K) into [Fe/H]}
\label{transf}

An early empirical relation between $V-K$ and [Fe/H] was proposed by 
Brodie \& Huchra (1990) based on Milky Way and M31 clusters, with the form: 
[Fe/H]$=1.574\cdot(V-K) -5.110$, the scatter around this relation being on the 
order of 0.27 dex. Clusters with metallicities from [Fe/H]$\simeq -2.3$
dex to [Fe/H]$\simeq -0.5$ dex ($V-K < 3.0$) were available to define
the relation.

A more recent calibration was proposed by Barmby et al.~(2000), based on
a newer compilation of Galactic clusters. They obtained shallower slopes
for the relation (1.30, if only Galactic clusters with low reddening
were used, 1.40 otherwise). Using their sample of M31 clusters 
they get a steeper slope with a relation of the form
[Fe/H]$\propto \sim1.90 \cdot (V-K)$. 

\begin{figure}[ht]
\psfig{figure=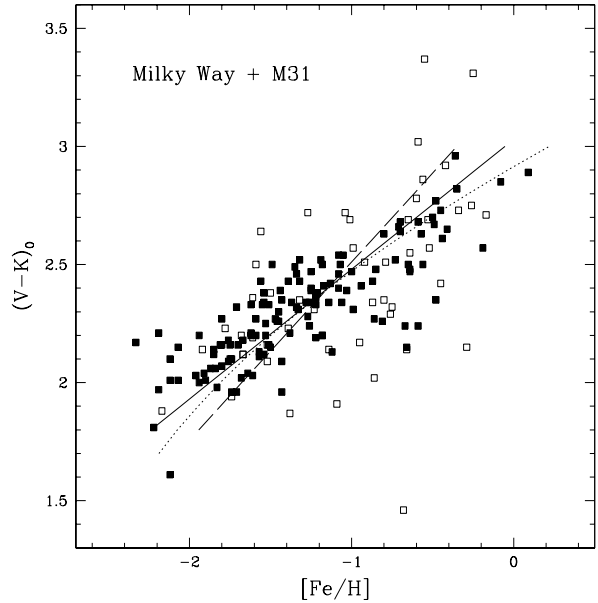,height=8cm,width=8cm
,bbllx=8mm,bblly=57mm,bburx=205mm,bbury=245mm}
\caption{ 
Relation between $(V-K)$ and [Fe/H] for globular clusters in the Milky
Way and M31. Solid points indicate clusters with a reddening
$E(B-V)<0.27$, open points clusters with higher reddening. Linear least squares
fits to the low-reddening and total sample are indicated as solid
and short-dashed lines, respectively. A second-order polynomial fit is
shown as dotted line.
}
\label{fig:relation}
\end{figure}

Here we use all available Galactic and M31 clusters from the above
compilations to derive a new relation. We consider only clusters with
low reddening (E$(B-V)<0.27$, including three quarters of the M31 data,
and 129 data points in total). We obtain a linear relation of the
form:

[Fe/H]$=1.82(\pm 0.11)\cdot (V-K)-5.52(\pm 0.26)$

(with an rms of 0.29 dex)

The data and relation are shown in Fig.~\ref{fig:relation}.
We also attempted to fit a second-order polynomial to the data and obtained
a best fit of the form:
[Fe/H]$=-2.33(\pm 1.81) - 0.92(\pm 1.55)\cdot(V-K) + 0.59(\pm
0.33)\cdot(V-K)^2$, with a standard deviation of the residuals of 0.29, a
reduced $\chi ^2$ of 0.0855 and an f-test result of 134, indicating a
poor fit. 

In the following, we will use our new linear relation, stressing that it
is only valid in the colour range $1.8<(V-K)<3.0$ and metallicity range
$-2.3<$[Fe/H]$<-0.2$ dex, and that it requires uncertain extrapolations outside
these ranges, especially towards higher metallicities. There is a
hint that the relation has a shallower slope at the high metallicity
end.  We will return to this point in Sect.~\ref{discu}.


\begin{figure}[ht]
\psfig{figure=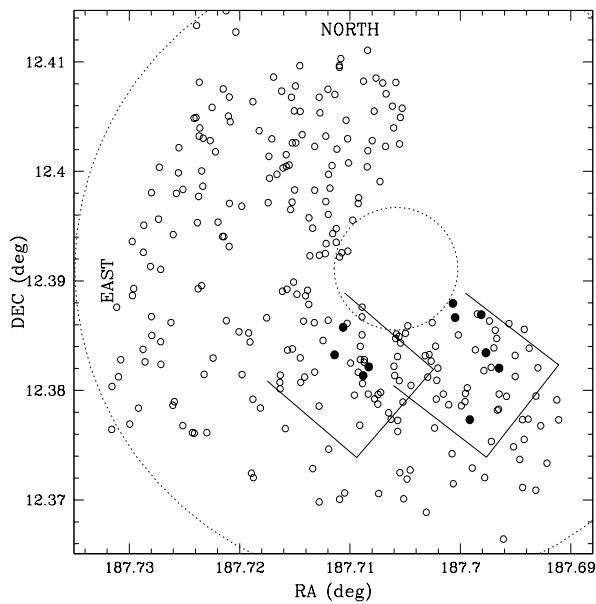,height=8cm,width=8cm
,bbllx=8mm,bblly=57mm,bburx=205mm,bbury=245mm}
\caption{ 
Position of objects with $V-K>3.4$ in M87 (solid dots) within our
$K$-band fields (3-sided frames) compared to the
position of all objects detect on the WFPC2 images with colours
$0.8<V-I<1.5$ and $21<V<23$ (open dots). The outer circle marks 1
R$_{\rm eff}$ (104\arcsec ), the inner one 20\arcsec . 
}
\label{fig:redpos}
\end {figure}


\subsection{In M87}

The colour distribution in M87 is broad, spanning over 2 magnitudes in
colour. Most globular cluster candidates span the range $1.8<V-K<3.4$ with 
several candidates having colours as extreme as $V-K\sim 1.5$ and $V-K\sim 4.3$.
As a comparison, the most metal-rich globular clusters in the Milky Way
and M31 have colours of $V-K<3.2$.

\subsubsection{Very red globular cluster candidates}

Figure \ref{fig:redpos} shows the location within our NIR fields of the
clusters with very red colours ($3.4<V-K<4.0$).  These objects were all
individually inspected by eye on the WFPC2 images, but none showed any
peculiarity that would disqualify it as a good globular cluster
candidate. They do not seem to be spatially clumped nor are they concentrated
towards the center of M87 (as verified by monte carlo simulations). 
Thus their morphological properties do not permit any useful conclusions
to be drawn concerning the nature of these objects.
Note that this group of clusters does
not stand out in $V-I$ since their {\it optical} colours are not exceptionally
red, and they blend in with the ``normal'' red tail of objects in the $V-I$
distribution. It is their $V-K$ colour that characterizes them.  A final
answer on their nature can only come from follow-up spectroscopy, for which
purpose we give in Table \ref{t:veryred} their magnitudes, colours and
positions (as returned by the task {\tt metric} within {\tt IRAF},
i.e.~relative positions probably good to 0.1\arcsec ).

\begin{table*}
\caption{Very red clusters in M87}
\label{t:veryred}
\begin{tabular}{l r r c c c}
\hline
\noalign{\smallskip}
ID & $\Delta$RA & $\Delta$DEC & $V_0$ [mag] & $(V-I)_0$ [mag] & $(V-K)_0$ [mag] \\
\noalign{\smallskip}
\hline
\noalign{\smallskip}
KPBMred1 &$ 16.8$ & $-19.3$ & $22.704\pm0.043$ & $1.094\pm0.061$ & $4.35 \pm 0.08$\\ 
KPBMred2 &$ 19.6$ & $-28.3$ & $23.600\pm0.051$ & $1.277\pm0.070$ & $3.59 \pm 0.16$\\ 
KPBMred3 &$ 10.5$ & $-35.1$ & $21.681\pm0.041$ & $1.202\pm0.058$ & $4.39 \pm 0.06$\\ 
KPBMred4 &$  8.7$ & $-32.3$ & $23.105\pm0.046$ & $1.147\pm0.065$ & $3.92 \pm 0.10$\\ 
KPBMred5 &$-32.8$ & $-11.4$ & $22.865\pm0.045$ & $1.185\pm0.063$ & $3.82 \pm 0.09$\\ 
KPBMred6 &$-28.7$ & $-32.7$ & $23.340\pm0.049$ & $1.242\pm0.068$ & $3.46 \pm 0.13$\\ 
KPBMred7 &$-27.1$ & $-27.6$ & $22.503\pm0.043$ & $1.115\pm0.060$ & $3.48 \pm 0.11$\\ 
KPBMred8 &$-18.8$ & $-15.1$ & $23.752\pm0.057$ & $1.240\pm0.079$ & $3.81 \pm 0.15$\\ 
KPBMred9 &$-13.3$ & $-16.1$ & $22.848\pm0.044$ & $1.210\pm0.062$ & $3.51 \pm 0.09$\\ 
KPBMred10&$-23.5$ & $-49.6$ & $22.929\pm0.045$ & $1.310\pm0.062$ & $3.82 \pm 0.11$\\ 
\noalign{\smallskip}
\hline
\end{tabular}
\begin{list}{}{}
\item[The positions are given relative to the centre of M87: 
12h30m49.4s $+$12d23m28s (J2000)]
\end{list}
\end{table*}

\subsubsection{From $(V-K)$ to [Fe/H]}

These colours can be translated into metallicity either using population
synthesis models, or using empirical calibrations derived e.g.~from the
Milky Way or M31 globular clusters (see last section). 

Using the empirical relation derived above, the globular clusters around M87 
would have metallicities ranging from [Fe/H]$=-2.7$ dex to 0.75 dex.
This broad range is supported by a comparison with population synthesis
models (see Fig.~\ref{fig:cmd}).

The histogram of colours for clusters in M87 clearly shows two major peaks.
This can be quantified by applying the KMM test (Ashman et al.~1993),
which tests the hypothesis that a single Gaussian is a better fit than the
sum of two Gaussians. The KMM test, run on our dataset selected for
clusters with $V-K<4.0$ (87 objects), returns a probably of $P=0.066$,
clearly favouring two Gaussians. The two colour peaks are found at
$V-K=2.11\pm 0.02$ and $V-K=3.13\pm 0.02$. 
These colour peaks translate into metallicity peaks of [Fe/H]$=-1.68$ 
dex using our relation derived in sect.~\ref{transf} ($-1.79$ dex
when using the Brodie \& Huchra relation) for the metal-poor clusters, 
and of [Fe/H]$=+0.18$ dex ($-0.18$ dex) for the metal-rich clusters 
(all values with total errors around 0.3 dex).

M87 has long been known to host (at least) two distinct populations
(e.g.~Lee \& Geisler 1993, Whitmore et al.~1995). The most recent study
by Larsen et al.~(2000) found the two peaks in $V-I$ at 0.95 and 1.20 mag
(within errors of a few hundreds of a magnitude), which translate into
metallicity peaks of [Fe/H]$\sim -1.4$ and $-0.6$ dex, when using the
relation given by Kissler-Patig et al.~(1998), or [Fe/H]$\sim -1.3$ and
$0.0$ dex, when adopting the relation from Couture et al.~(1990).
Thus, the combination of NIR and optical colours suggest a slightly more
metal-poor blue sub-population and a slightly more metal-rich red
sub-population in M87.


\subsection{In NGC 4478}

The colour distribution of globular clusters in NGC~4478 is very different from 
the one in its giant neighbour. About 18 globular cluster candidates were 
detected in our field, compared to a total population of $\sim 40$,
derived by Neilsen et al.~(1997). As in the optical study, hardly any
red globular clusters were detected, despite the fact that our observations
were more sensitive to red objects.

A KMM test for bi-modality on the small sample is inconclusive.
The mean $V-K$ colour of the sample is $2.31\pm0.03$ mag (and a dispersion
$\sigma = 0.60$ mag). This corresponds to a metallicity of
[Fe/H]$=-1.32$ dex using our relation of sect.~\ref{transf} ($-1.47$ dex
when using the Brodie \& Huchra relation).
Computing the mean for the sample bluer than $V-K=3.0$ (15 objects,
leaving out the 1 red cluster)
returns a mean colour of $V-K = 2.10\pm0.03$, with dispersion of 0.39 mag.
This corresponds to a metallicity of [Fe/H]$=-1.70$ dex ($-1.80$ dex).
Identical to the blue sub-population in M87.


\section{Discussion}
\label{discu}

\subsection{A comparison between M87, NGC 4478, and the Local Group systems}
\label{s:comp}

\begin{figure}[ht]
\psfig{figure=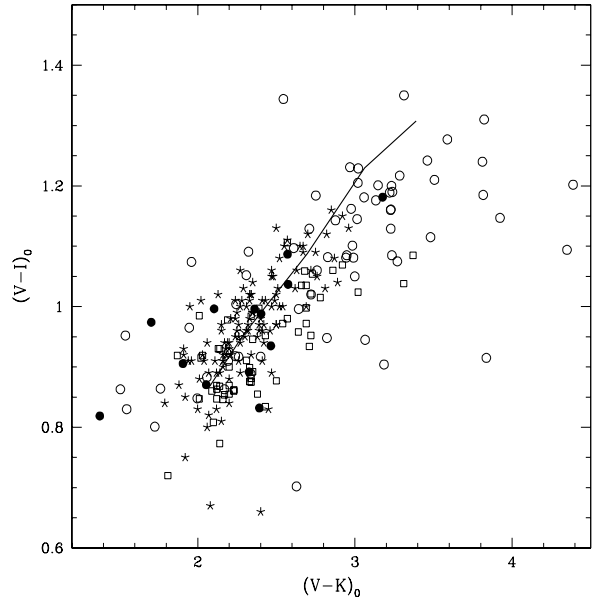,height=8cm,width=8cm
,bbllx=8mm,bblly=57mm,bburx=205mm,bbury=245mm}
\caption{ 
Colour-colour diagram showing globular clusters of M87 (open dots), NGC
4478 (solid dots), the Milky Way (open squares) and M31 (stars).
For comparison, a solid line shows a 15 Gyr isochrone
(from $1/200$ to twice solar metallicity) of a simple stellar population
(using here Maraston (2000) models).
}
\label{f:vivk_4}
\end {figure}


In the following, we compare the colour/metallicity distributions of M87
and NGC 4478 to the ones of the Milky Way and M31.
Figure \ref{f:vivk_4} shows a colour--colour plot for all four samples.
Note that the Local Group samples are biased towards metal-poor
clusters, with the most metal-rich ones having metallicities around
[Fe/H]$\sim-0.3$ dex.

For M87 the combination of optical and near-infrared colours showed 
more clearly than the optical colours alone, that the giant galaxy hosts
extreme clusters when compared to the Milky Way and M31. 
While all the clusters in the colour--colour diagram lie on the relationship
defined by the Local Group galaxies, M87 extends this relation to
extremes.  The blue
globular clusters extend to slightly bluer $V-K$ colours than the bulk of halo 
cluster in the Local Group, while the red clusters are on average redder 
in $V-K$ than even the reddest Local Group globular clusters.

The study of a large sample of galaxies showed that the populations of
metal-poor clusters appears to have a fairly universal average metallicity
around [Fe/H]$\sim -1.4\pm0.2$ (mostly derived from optical colours
alone: Burgarella, Kissler-Patig \& Buat 2001,
Larsen et al.~2001) with very little dispersion from galaxy to
galaxy. The wider
baseline of the $V-K$ colour suggests that M87 has, on average, very
metal-poor clusters in its metal-poor population (if colour is interpreted
solely in terms of metallicity).

The metal-rich clusters in our M87 sample are found to have around solar 
metallicities (see also fig.~\ref{fig:cmd}). $V-I$ studies already suggested 
this result, but are affected by transformation uncertainties
between colour and metallicity. In addition, the colour-metallicity
relation in the optical flattens at high metallicities which makes it even
more difficult to detect very metal-rich clusters in $V-I$. 
The $V-K$ colour allows a direct comparison with Milky Way and M31 globular 
clusters and clearly shows the redder colours of the M87 
clusters. This trend is stronger in $V-K$ than in $V-I$ which appears to
``saturate'' at high metallicities.

NGC 4478 hosts an almost exclusively blue population of globular
clusters (see above). However, when compared to the Milky Way and M31,
the population is very similar in $V-I$ and $V-K$ to the (halo) populations of
the Local Group galaxies. Also, the mean metallicity is in very good
agreement with the ``universal'' value found for metal-poor populations
in other early-type galaxies.

\subsection{A comparison with SSP models}
\label{s:young}

\begin{figure}[ht]
\psfig{figure=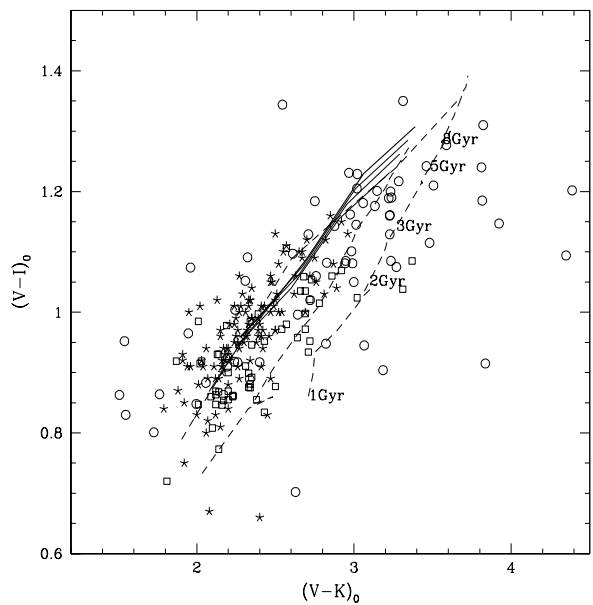,height=8cm,width=8cm
,bbllx=8mm,bblly=57mm,bburx=205mm,bbury=245mm}
\caption{ 
Colours of M87 globular clusters compared with models. The solid lines
show isochrones (from 1/200th to twice solar) of 15, 13, 11 and 9 Gyr
(models from Maraston 2000). The dashed lines show a 15 Gyr isochrone
as well as two iso-metallicity lines
(solar and twice solar from 1 to 15 Gyr) from Bruzual \& Charlot (2000).
Other symbols as in Fig.~\ref{f:vivk_4}.
}
\label{f:m87_models}
\end {figure}

The sample of globular clusters around M87 is well-suited to testing
population synthesis models in the near infrared. As seen in
fig.~\ref{f:vivk_4}, the Local Group simple stellar populations
(i.e.~globular clusters) do not provide samples extending to very red
($V-K>3.0$) colours.

Overall, the clusters are well-traced in the colour--colour plot by a 15
Gyr isochrone. However, for colours redder than $V-K>3.0$, the clusters
seem to deviate from the old models. The red clusters lie ``below'' the old 
isochrone (e.g.~see fig.~\ref{f:vivk}). 

In figure \ref{f:m87_models}, 
we show a set of younger isochrones (9 to 15 Gyr), but these also fail to 
match the reddest data points. Only when extending the isochrones to very high 
metallicities (twice solar) and allowing for younger ages (from 1 to 8 Gyr), the
models occupy the region populated by the reddest M87 clusters.
According to the models, these clusters would have solar
to twice solar metallicities and ages between 2 and 8 Gyr.

At these younger ages, a cluster would be expected to be 0.5 to 1 magnitude
brighter than an old (15 Gyr) cluster of the same mass and metallicity. This
luminosity difference is too small to allow us to use the apparent
magnitudes as age discriminators. We only notice that the reddest clusters
are among the {\it faintest} detected clusters, although still around the
visual turn-over magnitude of the globular cluster luminosity function. If
the reddest clusters have a mass distribution similar to the blue and red
ones, we can exclude the existence of a large number of them since we would
have detected many more at brighter magnitudes.

To quantify the fraction of intermediate-age clusters is
difficult. A simple attempt can be made by comparing the number counts of
clusters compatible with being old and metal-rich with
the number of intermediate-age candidates. The association of a cluster 
with the one or the other group is uncertain, and we try to bracket the
ratio with two extreme cases. For both cases, we base our counts on the data as
shown in Fig.~\ref{f:m87_models}, with $(V-K)_0>2.6$ in order to exclude
the old, metal-poor sub-population. In the first case, we consider
only clusters on the twice solar track with ages $<5$ Gyr to be of 
intermediate age. In this case, we obtain an old/young ratio of 4/1. In
the second case, we consider only clusters on $>9$ Gyr isochrones and up to
solar metallicity to be old and metal-rich. In that case, we obtain a ratio
old/young of 1/1. These ratios must be corrected for two biases: {\it i)}
the intermediate age clusters are on average redder and thus less
affected by incompleteness; and {\it ii)} the intermediate-age clusters
will be up to 1 mag brighter, i.e.~we sample a large fraction of the
total sub-population (assuming that old and intermediate age clusters
have the same mass function). These two effects amount to $10-100$\% 
correction, mostly depending on the exact age and mean colour of the
intermediate age clusters. The fraction of intermediate-age clusters in
our sample is thus estimated to be between 10\% and 50\%.
Finally, these number counts are made on a small area close to the center of 
the galaxy. Therefore statistical fluctuations as well as different spatial 
distributions can further change the ratios for the overall globular cluster 
system of M87.

Would clusters with such young ages have been detected in the
spectroscopic studies? Indeed, Brodie \& Huchra (1991),
Kissler-Patig et al.~(1998), Cohen et
al.~(1998) and Forbes et al.~(2002) have reported a few clusters in
early-type galaxies with
unusually high Balmer-line strengths for their metallicity, possibly
indicating young ages. These clusters were among the reddest of the
sample, although optical colours alone do not allow their identification as
extremely red. Also, the age-metallicity degeneracy, as well as
uncertainties in the structure of the horizontal branches prevent
definitive disentangling of age and metallicity effects using
spectroscopic line indices alone.

In summary, the metal-poor globular clusters are very well represented
by the models, which are very similar in the $(V-I)$ -- $(V-K)$ plane for a 
wide age range (8--15 Gyr) at low metallicities. 
The bulk of the metal-rich clusters ($V-K<3.0$) are also well represented by
the models and are compatible with old isochrones. Any systematic age
difference with respect to the metal-poor globular clusters is very
model dependent and is in any case smaller than a few Gyr.
For the reddest ($V-K>3.0$) globular clusters in M87, old isochrones can no longer  
fit the data, and the models seem to indicate the presence of very
metal-rich ($>solar$), intermediate age (2--8 Gyr) clusters. The colours 
for these clusters are comparable to optical-NIR colours of the confirmed 
intermediate age (few Gyr) clusters in NGC 1316 (Goudfrooij et al.~2001).
The very reddest ($V-K\sim4$) clusters remain a puzzle in terms of simple 
stellar population models, but resemble the objects in Cen A
suspected to be young-to-intermediate age star clusters (Minniti et 
al.~1996).

\subsection{Where are NGC 4478's red clusters?}

NGC 4478 appears to have a normal metal-poor globular cluster
sub-population but its lack of metal-rich clusters is surprising. Other 
early-type galaxies are known to be {\it dominated} by metal-poor globular 
clusters, but they also host a significant population of metal-rich
clusters. The origin of such metal-poor dominated distributions is still widely debated 
but is usually attributed to the ``mix'' of formation scenarios,
namely accretion, disk mergers, or in situ formation
(see Gebhardt \& Kissler-Patig 1999, Brodie, Larsen \& Kissler-Patig
2000). 
To date, only one other giant elliptical has been shown to host exclusively
metal-poor globular clusters: NGC 4874, the central cD galaxies in Coma
(see Harris et al.~2000). We will return to this case at the end of this
section.

The immediate question is whether NGC 4478 also qualifies as galaxy
with exclusively metal-poor globular clusters.
That is, does NGC 4478 host any red globular clusters at all? Our study shows 1
globular cluster redder than $V-K=2.6$ (which is roughly the limit for `halo' clusters in
the Milky Way and M31). The expected contamination
from M87 at this galactocentric distance, according to the work of McLaughlin, 
Harris \& 
Hanes (1994) is around 6 clusters per arcmin$^2$ down to $V=24$ mag (roughly 
our detection limit in the red). The ratio of blue to red clusters at
this large galactocentric distance is around 2/1 (estimated for NGC 4472 by 
Rhode \& Zepf 2001). Therefore, we expect 1 to 2 red clusters belonging
to M87 over our total area of $\sim 0.8$ arcmin$^2$ around NGC 4478. 
Thus, the one red globular cluster seen around NGC
4478 is fully compatible with contamination from M87. Similar
arguments can be made from the optical data alone (e.g.~Neilsen et
al.~1997). {\it NGC 4478 is likely to have no red globular clusters at
all.}	

NGC 4478 is a `normal' elliptical galaxy with respect to known scaling laws
(Mg--$\sigma$ relation, fundamental plane), and ranks among the dozen
brightest galaxies of the Virgo cluster. It stands out only as being
a likely companion of M87. The projected distance between the two galaxies
is 8.5\arcmin , i.e.~$\sim 40$ kpc (assuming a distance to both galaxies
of 16 Mpc), well inside the extended cD halo of M87 in projection
(e.g.~Carter \& Dixon 1978). NGC 4478 has been classified as
a compact elliptical (Prugniel et al.~1987) because of its lack of an
extended envelope, and the influence of M87 is evident from the
large twist of its outer isophotes (Peletier et al.~1990,
Michard 1985, Caon et al.~1990), and the complex nuclear
morphology (van den Bosch et al.~1994). Thus, NGC 4478 is likely to have
been tidally truncated by M87. Could this be the explanation for the absence 
of red clusters?

A priori, the red clusters are more concentrated than the blue ones, so
that truncation would be expected to affect, if anything, the blue
population preferentially. The specific frequency of NGC 4478 is very low
($S_N\sim 0.6$, Neilsen et al.~1997), which would be a consequence of the
outer parts (rich in clusters but poor in stars) having been
stripped. However, the complete absence of red clusters leads us to
speculate that the tidal truncation was effective {\it before} the
formation of the red clusters.  In a hierarchical formation scenario the
first small structures (responsible for the formation of blue clusters)
were able to form and assemble, but the remaining gas from this phase could
have been stripped before or during the formation of the red population and
associated stars, hindering their efficient formation. Such a scenario
implies a low fraction of old metal-rich stars with respect to old
metal-poor stars. This is supported by the somewhat low metallicity and
high H$_\beta$ of the galaxy's integrated light (Gonzales 1993) which can
be interpreted as the presence of a high fraction of old metal-poor stars
(Maraston \& Thomas 2000), matching the globular cluster
population. Interestingly, NGC 4478 has a low $\alpha$-element enrichment
as judged from the Mg$_2$/$<$Fe$>$ ratio, which hints at
star formation on long timescales rather than a rapid collapse.
This could in turn explain the high fraction of (metal-rich) stars
(i.e.~low S$_N$). NGC 4478 appears to have built up a metal-rich
stellar population without forming metal-rich globular clusters. This
may indicate a peculiar star-formation history/mechanism compared to
other gE galaxies.
A systematic study of the stellar and globular cluster populations of tidally
truncated galaxies would be of great interest for understanding the early
stages of
galaxy formation/assembly.

Getting back to the case of NGC 4874: the galaxy differs from NGC 4478
in the sense that instead of being influenced by a central galaxy, the
former dominates its environment, and tidal truncation by a
neighbour seems unlikely. However, Harris et al.~observe a similar
contradiction (maybe even more severe in their case); the galaxy built up 
a significant metal-rich stellar population
without forming any metal-rich clusters. Harris et al.~arrived at a similar
conclusion; a mechanism must have acted to suppress the effective
formation of massive globular clusters during the formation of the ``bulge''.
They actually argue that a mechanism that would prevent clusters more
massive than 10$^5 M_{\odot}$ from forming would also lead to the
observed properties. In our case the optical data sample
the globular cluster luminosity function almost completely, pushing down
the limit to a few 10$^4 M_{\odot}$ (see also Neilsen \& Tsvetanov
1997). Any further comparison between the two galaxy remains, however,
difficult given their very different natures and probably formation
histories (central cD vs.~truncated giant elliptical).

\subsection{The non-logarithmic metallicity distribution of M87}

Finally, we ask whether the red sub-population in M87 itself shows
sub-structure. In other words, whether the population of red
clusters in M87 was built up by several mechanisms/events, or whether
the bulk of the red globular clusters formed in a single event.

\begin{figure}[ht]
\psfig{figure=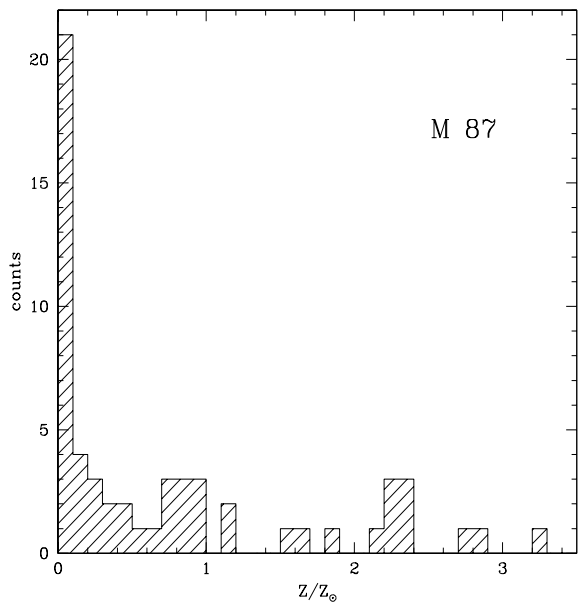,height=8cm,width=8cm
,bbllx=8mm,bblly=57mm,bburx=205mm,bbury=245mm}
\caption{ 
Metalicity distributions ($Z/Z_\odot$) of globular clusters in M87. 
Note that beyond $Z/Z_\odot=1$ the conversion from colour to metallicity
is based on an extrapolation of the empirical relations.
}
\label{f:zhistom87}
\end {figure}

The colour histogram (equivalent to [Fe/H] given the linear correspondence) 
shows the distribution of a logarithmic metallicity value. 
In Fig.~\ref{f:zhistom87} we show how different such metallicity distributions
would look in linear metallicity Z. In the case of M87, the two
clear peaks in [Fe/H] (transformed linearly from $V-I$) disappear when
plotted in Z. For this particular choice of zero point (solar$=$1), the blue 
peak gets ``compressed'' into an even clearer peak between 0 and
0.1--0.2 Z$_\odot$.
The (typical) gap in colour at [Fe/H]$\sim -1$ dex can still be guessed at
Z$\sim$0.1, since it marks the clear cut from the peaked to the roughly
flat distribution. The red peak gets spread over several tenths in Z
and is not recognizable as a single peak anymore. 

We conducted simple simulations to determine to what extent the
disappearance of the red peak was due to the fact that the 
(symmetric) photometric errors 
are asymmetrically distributed in the Z plot.
We simulated the M87 colour distribution using the peak location, peak
width, total numbers, and blue to red number ratios returned by the KMM
test. Each simulated data point was smeared by an error drawn randomly
from our observed list of errors in $V-K$.
The resulting 1000 artificial
distributions were compared to the original
data using a Kolmogorov-Smirnov test, in order to test whether a system
composed by 2 sub-population could reproduce the observed Z distribution.

We varied the peak locations, the peak width, and the number ratio of blue to
red clusters. We did this for a total of about 20,000 simulated distributions.
If the intrinsic widths of the sub-population was kept narrow (not
significant when compared to the photometric errors, here $\sigma
< 0.2$ mag)
the resulting distribution completely failed to reproduce the observed
one (the effect being more severe for the red ones). This excludes the possibility that
the globular cluster system of M87 is built from two sub-population of
very narrow metallicity range. This is certainly true for the red
sub-population, while this representation is not very sensitive to the 
intrinsic width of the blue sub-population in [Fe/H].
The best fit was obtained by allowing the red population to have a intrinsic
width of 0.25 to 0.30 mag in $V-K$ (i.e.~$\sigma\sim 0.5$ dex in
[Fe/H]), although even in that case the confidence level of the test
was only around 80\%, on average. 
Unfortunately, our number statistics were not high enough to probe more
complicated metallicity distributions, or 3 sub-populations. But we
conclude from these data, that the red distribution has a spread in
metallicity of $\sim$0.5 dex (1$\sigma$) roughly from half-solar to twice
solar, or even wider if age affects the
colours too (see Sect.~4.2). 
This spread in metallicity gets ``played down'' in the colour
and [Fe/H] distributions.

Thus, the interpretation of the metal-rich peak appears more
complicated than originally assumed and the red population might well
host multiple sub-populations.
The large spread in $V-K$ is naturally explained by several successive
events of red cluster formation, induced by successive gas-rich mergers
as assumed in hierarchical formation scenarios. These dissipational
events must have occurred at high redshifts (given that most clusters
appear to be old), but a small fraction of them could have happened more 
recently ($z\sim0.5$--1) given the potential presence of intermediate age 
(few Gyr) clusters (see Sect.~\ref{s:young}). The bulk of the
star/cluster formation in M87 would nevertheless have occurred at high 
redshifts.

Alternatively, the large spread in $V-K$, i.e.~in metallicity, is also
compatible with a narrow age distribution, as expected if the
metal-rich stellar and cluster populations had formed 
in a single event ({\it in situ} collapse or one single major
merger).  This would require extremely fast chemical enrichment during
this major star/cluster formation event, so that almost coeval clusters
could have very different metallicities. 
The presence of intermediate age clusters contradicts this
hypothesis, i.e.~the formation of the metal-rich component in a single
collapse. The current data favour a more complex star/cluster
formation history for M87 with several major star formation events
having played a role from high to intermediate redshifts.


\section{Summary and Conclusions}

We have compared optical and near infrared colours of globular
clusters in M87, the central giant elliptical in Virgo, and NGC 4478,
an intermediate luminosity galaxy in Virgo, close in projection to
M87.

We find the broad range
in colour and previously detected bi-modality in M87. We confirm that NGC 4478
only hosts a blue sub-population of globular clusters and now show that
these clusters' $V-I$ and $V-K$ colours are very similar to those of the 
{\it halo} globular clusters in
Milky Way and M31. Most likely, a metal-rich sub-population never formed
around this galaxy (rather than having formed and been destroyed later),
perhaps because its metal-rich gas was stripped during its passage through the
centre of the Virgo cluster,
or because it was subjected to some other mechanism that would prevent the formation of
massive globular clusters during the formation of the ``bulge''.

In M87 the $V-I$/$V-K$ colours are close to those predicted from 
SSP models for old populations. However, M87 also hosts a few red clusters 
that are best explained by intermediate ages (few Gyr). Generally, there is 
evidence that the red, metal-rich sub-population has a complex structure and is
itself composed of clusters spanning a large metallicity (from the spread
in $V-K$) and, potentially, age range (from the spread in the $V-I$--$V-K$
diagram). This contrasts with the blue, metal-poor population which appears 
more homogeneous from galaxy to galaxy.

While the blue clusters still appear to be predominantly a single physical
group, this may not be true for the red
clusters. This prompts us to ask whether the metal-rich
sub-group is, in many or most cases, just an amalgamation of multiple
metal-rich sub-populations that do not necessarily share the same origin.
This would imply a complex star/cluster formation history, i.e.~the origin of
the metal-rich stellar population in early-type galaxies may not be just a 
single collapse or a single major merger but rather several such events.

\begin{acknowledgements}

We would like to thank Claudia Maraston and Stephane Charlot for
providing their population synthesis models prior to publication.

MKP gratefully acknowledges the support of the Alexander von Humboldt
Foundation through a Feodor Lynen Fellowship in the early stages of this
project. 
DM is supported by Fondecyt, Fonfap, and Milenio, and he
also thanks the ESO Visitor's program.
Part of this research was funded by the National Science
Foundation grant number AST-9900732 and HST grant GO.06554.01-95A.

The authors wish to extend special thanks to those of Hawaiian ancestry on 
whose sacred mountain we are privileged to be guests. Without their generous 
hospitality, none of the observations presented herein would have been
possible.

\end{acknowledgements}


\end{document}